# Flux dependence of helium retention in clean W(110): experimental evidence for He self-trapping

A. Dunand, M. Minissale, T. Angot and R. Bisson

*Aix-Marseille Univ, CNRS, PIIM, UMR 7345, Marseille F-13397, France*

Corresponding author: regis.bisson@univ-amu.fr

**Abstract**

Helium (He) retention in tungsten (W) is a concern in fusion reactors since it could be detrimental to plasma facing components performance and influence the fusion fuel balance. He being not soluble in W, it tends to agglomerate on preexisting defects (vacancy, grain boundary), but it could in theory also self-trap (be immobilized on a non-preexisting vacancy) through the emission of a vacancy/self-interstitial W pair in the vicinity of a $He_n$ interstitial cluster. In the present study, we prepared a pure single crystal W(110) sample with a clean surface in order to evidence the self-trapping of He in the W bulk at a sample temperature of 300 K and for a constant fluence of $2.0 \times 10^{21}$ $He^+.m^{-2}$. At a $He^+$ kinetic energy of 130 eV and a flux of $0.3 \times 10^{17}$ $He^+.m^{-2}.s^{-1}$, we only observed a small He desorption peak below 600 K. Rising the ion flux to $0.7 \times 10^{17}$ $He^+.m^{-2}.s^{-1}$, we observed the sudden appearance of two desorption peaks at 950 K and 1700 K. For the highest flux studied in this work, $5.0 \times 10^{17}$ $He^+.m^{-2}.s^{-1}$, an additional desorption peak at 1800 K and a desorption shoulder at 1900 K are observed. The temperature position of these He desorption peaks are consistent with the density functional theory literature and points to the occurrence of self-trapping once the $0.7 \times 10^{17}$ $He^+.m^{-2}.s^{-1}$ flux is attained at 300 K and to the possible realization of trap-mutation for the flux of $5.0 \times 10^{17}$ $He^+.m^{-2}.s^{-1}$. The present set of results should be used to constrain the development of He retention and He bubbles growth models based on *ab initio* quantities.

## 1. Introduction

Fusion fuel retention and release from plasma facing components are critical issues for ITER and for any future prototype reactors such as DEMO. Thus, understanding fundamental mechanisms behind the retention of hydrogen isotopes in first wall and divertor materials is necessary. In a tokamak operating with a burning plasma, the divertor is subjected to a significant flux of helium ions ($He^+$). Additionally, in the current planning of the ITER operational phases, He plasma are planned early on to demonstrate high confinement mode without nuclear activation of the vacuum vessel [1]. Since, it has been demonstrated that hydrogen isotopes retention is strongly influenced by the presence of He in W divertor materials [2–4], understanding the processes leading to He retention in tungsten (W) is also of importance. He being not soluble in W it tends to agglomerate leading to the appearance of He-filled nanovoids (bubbles) [5]. These He bubbles are currently thought to be precursors to the formation of W "fuzz" [6–8], a nanostructured W materials with large surface area and degraded thermal conductivity which may be detrimental to the W material lifetime [9].

Most of models for He retention consider that He is a mobile species that agglomerate on preexisting defects, such as grain boundaries, vacancies and vacancy clusters. Grain boundaries have been observed to be decorated with He bubbles [10] while vacancies are thought to be responsible for He retention in the bulk [11], consistent with an increased retention when He$^+$ implantation kinetic energy is above ~500 eV [12,13], the knock-on threshold for displacement damage induced by He in W. However, another mechanism is possible in theory, He self-trapping [14], when for sufficient He$^+$ implantation flux there is formation of mobile interstitial He clusters (He$_n$) in W that can be immobilized on a "self-created" vacancy thanks to the emission in its vicinity of a Frenkel-pair (V+SIA, where V is a W vacancy and SIA is a self-interstitial W atom). The self-trapping mechanism can be represented by the equation He$_n$ → He$_n$V + SIA. Such self-trapping mechanism of He in W has been considered in several modeling works [15–19] but there is currently no definitive experimental proof that He can self-trap in W.

Previous experimental studies by Kornelsen [12,20] showed that 250 eV He$^+$ implantation in single crystals W(100) and W(110) at room temperature and in the $10^{18}$ He$^+$.m$^{-2}$ fluence range lead to the quasi absence of He retention, unless bulk vacancies are preexistent in significant quantities because of keV rare gas pre-implantation. In presence of such concentrations of preexistent vacancies and vacancy clusters (V$_n$), Kornelsen observed temperature programmed desorption (TPD) peaks linked to He retention at 950 K, 1120 K, 1220 K, 1560 K and 1675 K. Based on their fluence dependencies, these peaks were assigned to, respectively, the following dissociation pathways He$_4$V→He + He$_3$V, He$_3$V→He + He$_2$V, He$_2$V→He + HeV, HeV → He + V and He$_2$V$_2$→2He + 2V. More recently, Iwakiri *et al.* [21] showed with *in situ* transmission electron microscopy (TEM) that 250 eV He$^+$ implantation into a W polycrystalline sample at 293 K resulted in the appearance of He clusters (platelets) in the bulk of W for a fluence above $1.4 \times 10^{19}$ He$^+$.m$^{-2}$, followed by emission of interstitial dislocation loops for a fluence of $3.0 \times 10^{19}$ He$^+$.m$^{-2}$. Iwakiri *et al.* proposed that He aggregation occurred around a bulk impurity which ejection frees a vacancy that immobilizes the He cluster, i.e. an impurity-assisted self-trapping mechanism. Note that Kornelsen highlighted the effect of surface impurities on He retention, since a He desorption peak appeared below 800 K if He implantation was realized on a surface left in vacuum for 48 hours without cleaning [12]. This surface effect may be related to the observation made by Gasparyan *et al.* that 3 keV He implanted polycrystalline W at 1000 K exhibited broad He desorption peaks below 1000 K after air exposure [22]. Notwithstanding these surface effects, we note two main differences between Kornelsen and Iwakiri *et al.* studies. First, the He$^+$ fluence where Iwakiri *et al.* observed the appearance of He clusters in the bulk was larger by an order of magnitude ($10^{19}$ vs $10^{18}$ He$^+$.m$^{-2}$) than the fluence used by Kornelsen for concluding to the absence of He self-trapping in bulk W. Second, the He$^+$ flux used by Iwakiri *et al.* was three order of magnitude higher ($10^{18}$ He$^+$.m$^{-2}$.s$^{-1}$ [23]) than the one used in Kornelsen's study ($5.10^{14}$ He$^+$.m$^{-2}$.s$^{-1}$ [12]). Modeling works about He self-trapping by Faney *et al.* [18] and Delaporte-Mathurin *et al.* [19] have highlighted that an increase of the He$^+$ flux leads to an increase of He retention in W because of the induced increased density of He$_n$ clusters with n→6 which makes the He self-trapping mechanism He$_n$ → He$_n$V a more probable event [17].

In this work, we aimed to test the results of modeling studies showing that the He$^+$ flux is a driving parameter for the self-trapping of He in the W bulk. Taking the studies of Kornelsen and Iwakiri *et al.* into account, we implanted 130 eV He$^+$ in a single crystal W with a clean surface (to avoid surface impurity effects) at 300 K and for a fluence above $3.0 \times 10^{19}$ He$^+$.m$^{-2}$. In these conditions deemed optimal to observe He self-trapping in the W bulk, we systematically varied the He$^+$ flux in the $10^{16} - 10^{17}$ He$^+$.m$^{-2}$.s$^{-1}$ range and determined a He$^+$ flux threshold on He retention as shown by TPD. In section 2, we detail the preparation of the W single crystal, the He$^+$ implantation as well as the TPD evaluation of He retention. In section 3, we present the effect of He$^+$ flux on the He retention measured by TPD and discuss the observed results. In section 4, we conclude on this work and propose future experimental studies to better constrain modeling developments.

## 2. Material and methods

The following experiments have been carried out in the Advanced MUltibeam experiment for Plasma Surface Interaction (AMU-PSI), an ultra-high vacuum (UHV) setup located at Aix-Marseille University (Marseille, France) [24]. We used two of the multiple interconnected chambers, the sample chamber (base pressure P ≈ $2 \times 10^{-8}$ Pa) and the ion beam/mass spectrometer chamber (base pressure P ≈ $5 \times 10^{-8}$ Pa), allowing the realization of TPD measurements after He$^+$ implantation without air exposure (termed *in situ* in the following). The W sample was mounted in the sample chamber on a molybdenum plate attached to a hollow OFHC copper support connected on a four-axis manipulator. Flowing compressed air in the OFHC copper support allowed to accelerate sample cooling to room temperature. The sample temperature was measured optically on the sample's back face with two 1-color pyrometers (SensorTherm METIS M323 and M313) interfaced through a dual-pyrometer controller (SensorTherm Regulus) allowing automatic switching of the measuring range to achieve a continuous 323 – 2173 K temperature reading. The emissivity was set to 0.3, which results in uncertainties on temperature measurement of about ±20 K according to our previous measurement of W reflectivity [25,26]. The W temperature was controlled by a computer-based PID regulator. The sample was a cylindrical (8 mm diameter, 2 mm thick) pure (99.999%) W single crystal (Surface Preparation Laboratory) oriented along the (110) crystallographic plane, aligned and mechanically polished with an accuracy of 0.1°. The W(110) sample was heated on the back face with a CW ytterbium fiber laser (SPI laser Qube 1000) delivering up to 1000 W at ∼1075 nm. Laser heating was used both to remove impurities and defects from the sample and for TPD measurements. A four-grid low-energy electron diffractometer (LEED)/Auger electron spectrometer (AES) (OCI BDL600IR) complemented the sample chamber and was used to characterize the crystalline structure and the chemical composition of the sample surface at room temperature. In the ion beam/mass spectrometer chamber, a commercial ion source (Focus FDG15) was used to produce a He$^+$ ion beam with a kinetic energy of 130 eV/He$^+$ or 620 eV/He$^+$. A leak valve on the ion beam inlet allowed to set precisely the He$^+$ ion flux in the $10^{16} - 10^{17}$ He.m$^{-2}$ s$^{-1}$ range as inferred from the measure of the ionic current on the sample [27]. The He$^+$ ion beam impinged the W(110) sample along the normal to its surface. We quantified the release of He atoms from the sample during TPD using a quadrupole mass spectrometer (QMS) (MKS Microvision2). A unique W(110) sample was used for all presented measurements.

To avoid preexistent bulk defects and surface impurities effects on He retention [12], the W(110) sample was thoroughly cleaned until LEED showed a sharp 1x1 diffractogram, consistent with the absence of carbon and oxygen impurities as measured by AES [27]. In addition, before every He$^+$ implantation/TPD experiment, the sample was annealed 10 times with a temperature ramp of 12.5 K.s$^{-1}$ up to 2173 K in order to remove vacancies created by self-trapping of He in former experiments.

He$^+$ implantation was then performed on W(110) at room temperature (300 K). The desired He$^+$ fluence was realized by time integration of the measured ionic current on the sample. After implantation, a waiting time of 2 hours allowed a sufficient reduction of the ion beam/mass spectrometer chamber He partial pressure such that we could perform the TPD measurement. After each TPD measurement, a second TPD measurement was realized 5 min after the first one. This second TPD was carried out to subtract it from the first one, i.e. to perform a background subtraction, since a recurrent and remote background source of D$_2$ (same mass than He for our QMS resolution) was remanent from the previous D$_2$ experimental campaign [27], especially at surface temperature above 2000 K. The origin of this D$_2$ background was traced back to the molybdenum plate on which the W(110) is attached as the background became more intense with increasing the molybdenum plate temperature (checked by thermocouple measurements on the plate). Therefore, some background subtraction lead to negative desorption rate at the highest temperature (> 2000 K) since between two subsequent TPDs the molybdenum plate temperature increased and its related D$_2$ background desorption increased. In the following, we present background-subtracted TPD measurements up to 2000 K. Note that we realized also neutral He gas exposure of W(110) to evaluate the residual signal from our background subtraction methodology.

TPD measurements were realized with a temperature ramp of 12.5 K.s$^{-1}$. TPD curves for each flux and fluence conditions were obtained after binning in 10 K intervals the QMS signal of individual TPD measurements and averaging between replicate TPD measurements. Note that the ion implantation was performed at 300 K but the pyrometer-based temperature regulation of the TPD was effective only from 413 K. Thus, from 300 K to ~450 K, TPD curves displayed a sharp-rising and slow-decreasing desorption peak because the TPD temperature ramp was not yet linear with a rise higher than 12.5 K.s$^-$ [1]. For this reason, we stress that the desorption rate in the TPD curves below 450 K are not realized at a constant temperature ramp of 12.5 K.s$^{-1}$. Nevertheless, the whole area under the raw TPD curve (i.e. as a function of time) can be used to evaluate the He retention.

He retention for each flux/fluence condition was evaluated by time-integration of the TPD curves. The absolute calibration of these time integrated curves in units of He.m$^{-2}$ was realized according to our previous study on $D_2$ dissociative adsorption and $D_2^+$ ion implantation on clean W(110) [27]. Saturated dissociative adsorption of $D_2$ on W(110) is known to give a deuterium retention of ~$1.0 \times 10^{19}$ D.m$^{-2}$ at 300 K [28]. Electron-impact partial ionization cross sections [29,30] were used to correct for the QMS sensitivity to He and $D_2$.

## 3. Results and discussion

### 3.1. He$^+$ flux dependency of He retention at a constant fluence

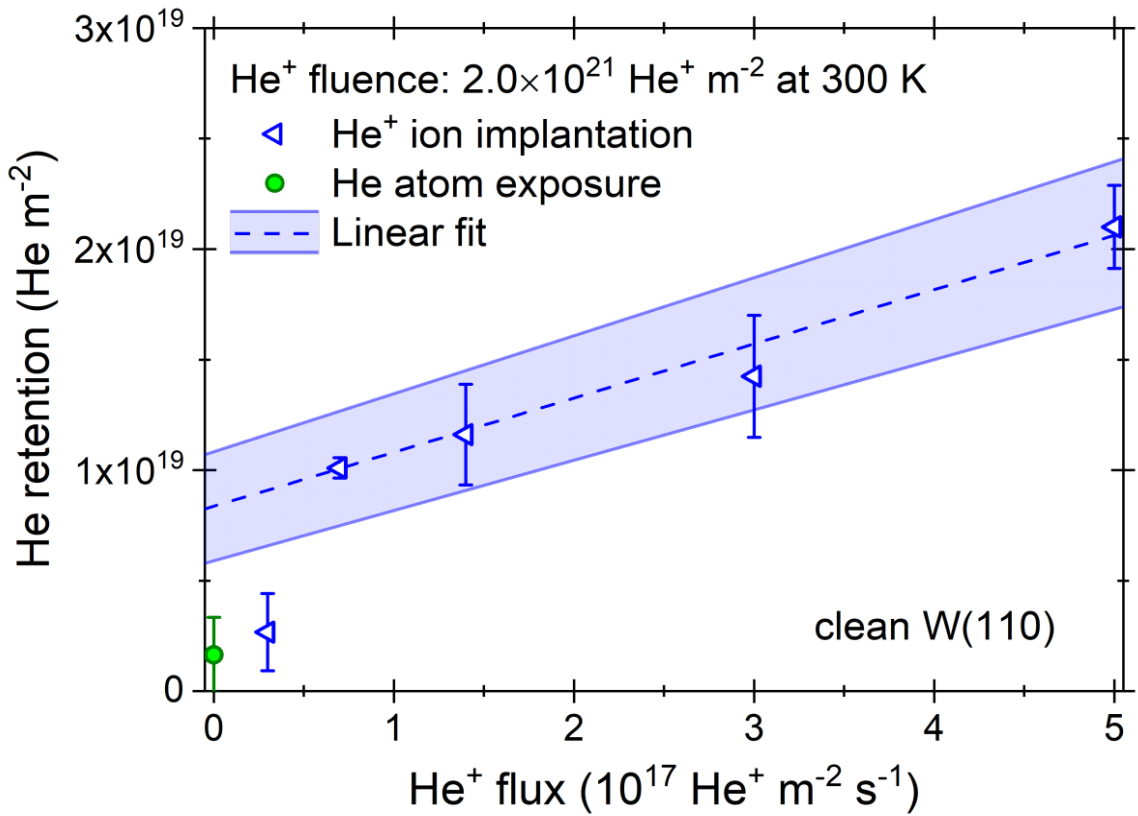

Figure 1. He retention as a function of He$^+$ flux (open blue triangle) measured *in situ* two hours after the implantation of 130 eV He ions in clean W(110) at a surface temperature of 300 K with a fluence of $2.0 \times 10^{21}$ He$^+$.m$^{-2}$. The neutral He gas exposure (green circle) represents the residual background obtained with our methods. Error bars are the standard deviation of the mean of replicate measurements. A linear fit (dashed line) realized on He$^+$ flux equal or above $0.7 \times 10^{17}$ He$^+$. m$^{-2}$.s$^{-1}$ is shown with its uncertainties (shaded interval) to highlight the He$^+$ flux threshold necessary for He retention through self-trapping.

Figure 1 shows the evolution of He retention in clean W(110) as a function of $He^+$ flux, for an identical ion fluence of $2.0 \times 10^{21}$ $He^+.m^{-2}$ impinging with a kinetic energy of 130 eV and at a surface temperature of 300 K. He retention after He gas exposure is also shown to inform on the residual background from the methods described in Section 2. This He gas exposure was realized through the ion gun turned off. For the lowest $He^+$ ion flux investigated here ($0.3 \times 10^{17}$ $He^+.m^{-2}.s^{-1}$), the obtained He retention is on par with the He gas exposure residual background. When increasing the $He^+$ flux by a factor of 2.3 (i.e. $0.7 \times 10^{17}$ $He^+.m^{-2}.s^{-1}$), the He retention is 6 times higher than the residual background and clear TPD desorption peaks are suddenly appearing (Figure 2). Then, above $0.7 \times 10^{17}$ $He^+.m^{-2}.s^{-1}$ the He retention is increasing linearly with the $He^+$ flux. This flux threshold is clearly seen in Figure 1 when fitting a linear law through the experiments where the flux is sufficient to bring about the emergence of TPD desorption peaks.

The observation of this flux threshold suggests that for $0.7 \times 10^{17}$ $He^+.m^{-2}.s^{-1}$ at 300 K the instantaneous density of He in W is sufficient to create $He_n$ interstitial clusters with n≥6, the minimum cluster size for which self-trapping occurs in bulk W according to Boisse *et al.* [17].

### 3.2. $He^+$ flux dependency of TPD He desorption peaks

Figure 2 presents the TPD measurements obtained from the clean W(110) sample following neutral He gas exposure or $He^+$ ion implantation at 300 K at various flux in the $10^{16}$-$10^{17}$ $He^+.m^{-2}.s^{-1}$ range. The neutral He gas exposure at 300 K leads to an absence of any He desorption peak, consistent with the know physisorption of He on W(110) [31]. At the lowest $He^+$ ion flux of $0.3 \times 10^{17}$ $He^+.m^{-2}.s^{-1}$, there is no He desorption peak above 600 K where self-trapped $He_nV_m$ clusters are expected (black and grey solid stars) according to numerous density functional theory (DFT) studies [17,32]. However, we observed a small He desorption peak below 500 K which extends up to ~ 600 K and could be the signature of interstitial $He_n$ clusters in bulk W as expected (open stars) from the DFT study of Boisse *et al.* [17]. This He desorption below 600 K rises in intensity as the $He^+$ flux is increased, indeed. An analysis using the method of Redhead [33], assuming a first-order desorption, using a $10^{13}$ $s^{-1}$ prefactor and our experimental temperature ramp of 12.5 $K.s^{-1}$, allows to transform DFT calculations by Boisse *et al.* for the dissociation energy of $He_n$ interstitial clusters into temperature of desorption peaks (open star in Figure 2). This Redhead analysis suggests that the He desorption below 600 K would correspond to $He_n$ interstitial clusters with n≤3.

Figure 2 shows also that as soon as the $He^+$ flux reaches $0.7 \times 10^{17}$ $He^+.m^{-2}.s^{-1}$, two desorption peaks appear, centered at 950 K and 1700 K, superimposed to a board desorption signal which is above the residual background and extends from 600 K up to 1500 K. For the highest flux of this study, $5.10^{17}$ $He^+.m^{-2}.s^{-1}$, the two desorption peaks increased in height while the second desorption peak shifts from 1700 K down to 1650 K. A third desorption peak centered at 1800 K is revealed as well as a desorption shoulder at 1900 K. Note that the broad desorption signal from 600 K to 1500 K is also further increased.

The measured He desorption peak centered at 950 K is consistent, through the Redhead analysis, with the DFT literature (black stars in Figure 2) for the dissociation energy of self-trapped $He_nV$ clusters, with a He content $n \geq 2$. The He desorption peak centered at 1700 K is also consistent with the DFT literature for the mean dissociation energy of 4.7 eV for $He_1V$. Thus, the TPD measurement just above the $He^+$ flux threshold exhibits a pair of broad desorption peaks consistent with immobilized $He_nV$ self-trapped clusters while below the $He^+$ flux threshold the only desorption peak is typical of mobile $He_n$ interstitial clusters. This is an evidence for the effect of the $He^+$ flux on the self-trapping of $He_n$ interstitial clusters in the W bulk.

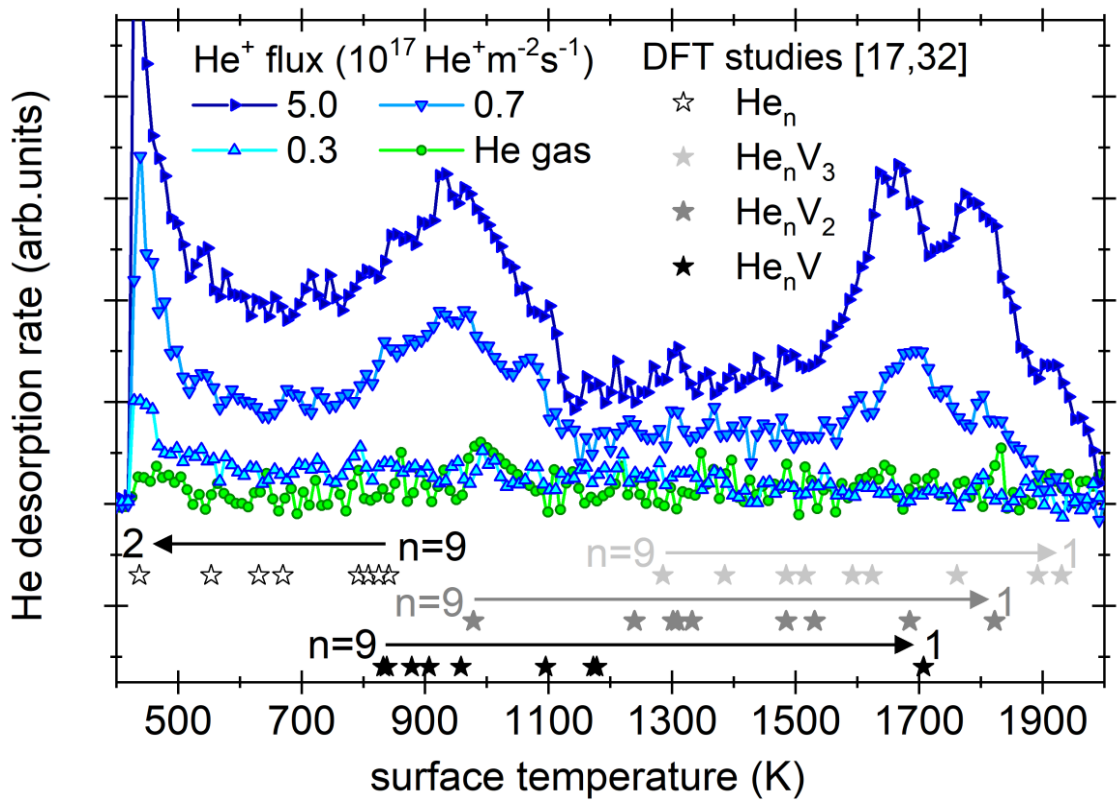

Figure 2. *in situ* TPD measured after the implantation of 130 eV He ions in clean W(110) at a surface temperature of 300 K with a fluence of $2.0 \times 10^{21}$ $He^+.m^{-2}$. The neutral He gas exposure (green circles) highlight the residual background obtained with our methods. Three different $He^+$ flux are shown: $0.3 \times 10^{17}$ $He^+.m^{-2}.s^{-1}$ (cyan up-triangle), $0.7 \times 10^{17}$ $He^+.m^{-2}.s^{-1}$ (blue down-triangle) and $5.0 \times 10^{17}$ $He^+.m^{-2}.s^{-1}$ (navy right-triangle). Desorption peak positions predicted using a Redhead analysis [33] for He dissociation energies from DFT studies [17,32] are presented for mobile interstitial $He_n$ clusters (open star) and immobilized $He_nV$ (black solid star), $He_nV_2$ (dark gray solid star) and $He_nV_3$ (light gray solid star) clusters.

Additionally, for the highest flux investigated here the third He desorption peak centered at 1800 K is consistent with the DFT literature for the $He_1V_2$ cluster (dark grey star) while the 1900 K shoulder corresponds to the DFT dissociation energy of the $He_2V_3$ and $He_1V_3$ clusters (light gray star). These vacancy clusters containing He are expected to occur for sufficient He density where the trap mutation mechanism $He_nV_m \rightarrow He_nV_{m+1} + SIA$ will further stabilize immobilized clusters that contains numerous He atoms [17]. Such vacancy clusters containing He can also explain the shift of the 1700 K desorption peak to 1650 K, thanks to $He_2V_2$ (dark gray star), $He_5V_3$ and $He_4V_3$ (light gray stars) clusters, and the observation of the increasing broad desorption from 600 K to 1500 K with increasing the flux thanks to $He_nV_m$ clusters for n≥4 and m=1, 2, 3 (solid stars between ~800 K and 1600 K in Figure 2).

Note that we measured three main desorption peaks at 950 K, 1700 K and 1800 K in our clean W(110) sample, while Kornelsen measured five main desorption peaks at 950 K, 1120 K, 1220 K, 1560 K and 1675 K in heavy-ion pre-implanted W(100) and W(110) clean samples [20]. Considering the little differences in the two *in situ* TPD setups, the absence of the two peaks at 1120 K and 1220 K in our experiments is striking. Actually, a He desorption peak around 1100 K is also seen in the TPD performed by Gasparyan *et al.* on polycrystalline W [22]. To better understand the absence of a desorption peak in the 1100 - 1200 K range in our sample, we realized a $2.0 \times 10^{21}$ $He^+.m^{-2}$ implantation at $5.10^{17}$ $He^+.m^{-2}.s^{-1}$ with a kinetic energy of 620 eV, i.e. above the knock-on threshold for displacement damage. In these conditions, a preponderant and broad He desorption peak centered at ~ 1100 K appears, that is 5 times more intense than all the other peaks (Figure 3). Thus, we conclude that pre-existing defects are responsible for a He desorption peak at ~1100 K, which could act as much efficient seeds, as compared to self-trapping in our flux conditions, for the growth of vacancy clusters containing He, i.e. $He_nV_m$ with m≥1 and/or large n/m ratio. Additionally, we observed a new and small desorption peak at 640 K upon the implantation of 620 eV $He^+$ ions at room temperature. This desorption temperature is reminiscent to the one observed by Kornelsen *et al.* [12]. and Gasparyan *et al.* [22] from surface contamination. However, in our present experimental setup surface contamination can be excluded. Instead, the appearance of this peak could be related to the growth of $He_nV_m$ clusters with large n/m ratio in the vicinity of the surface. Boisse *et al.* have shown with DFT that the He dissociation energy decreases from 5 eV down to ~2 eV as the n/m ratio increases to n/m~20 [17], indeed. Furthermore, Yang *et al.* have shown also with DFT that the He dissociation energy tends to diminish as $He_nV_m$ clusters get closer to the surface [34]. Taken together, these two DFT studies might explain the 640 K desorption peak observed presently, which could be translated through the usual Redhead analysis to a dissociation energy of ~1.8 eV for near-surface $He_nV_m$ clusters.

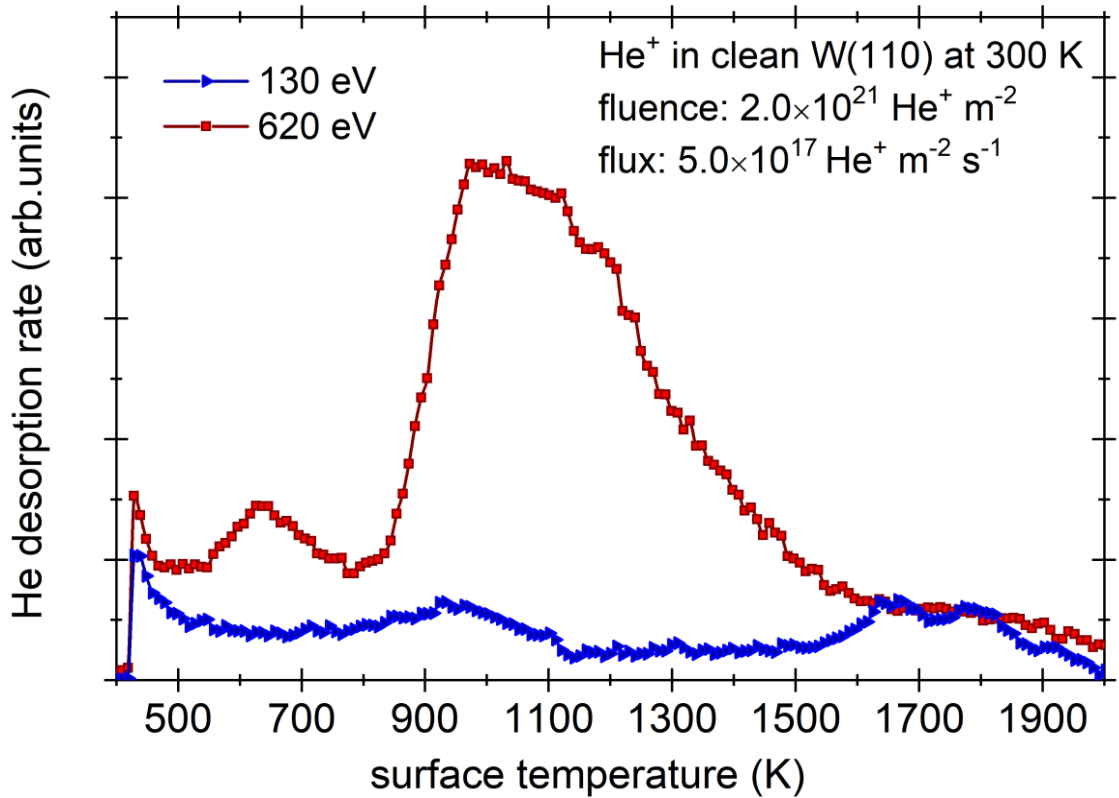

Figure 3. *in situ* TPD measured after the implantation of 130 eV (blue triangles) or 620 eV (red square) He ions in clean W(110) at a surface temperature of 300 K with a fluence of $2.0 \times 10^{21}$ He$^+$.m$^{-2}$ and with a flux of $5.0 \times 10^{17}$ He$^+$.m$^{-2}$.s$^{-1}$.

### 3.3. Comparison of He retention in single crystal and polycrystalline W

Figure 4 compiles the evolution of He retention as a function of He$^+$ fluence for two different studies where He$^+$ implantation is performed at room temperature, in a similar flux range and with kinetic energies below the knock-on threshold: the present results on clean W(110) and the recent work published by Gasparyan *et al.* on polycrystalline W [22]. For the same flux of $0.3 \times 10^{17}$ He$^+$.m$^{-2}$.s$^{-1}$ and the same fluence of $2.0 \times 10^{21}$ He$^+$.m$^{-2}$, the difference in He retention is two orders of magnitude, with our He retention being the lowest. We attribute this striking difference to a two-fold origin. First, in our experiment with a clean W(110) sample, the flux of $0.3 \times 10^{17}$ He$^+$.m$^{-2}$.s$^{-1}$ is below the threshold necessary for self-trapping thus only a thousandth of implanted He are retained two hours after the implantation. Second, even though self-trapping should be forbidden in Gasparyan *et al.*'s study, grain boundaries are present in their polycrystalline W as well as more impurities (about 50 times more in weight in their 99.95% sample). Therefore, we attribute the lowest He retention in our study to the fact that pre-existing defects and impurities are at their lowest level, so we are sensitive enough to self-trapping effects.

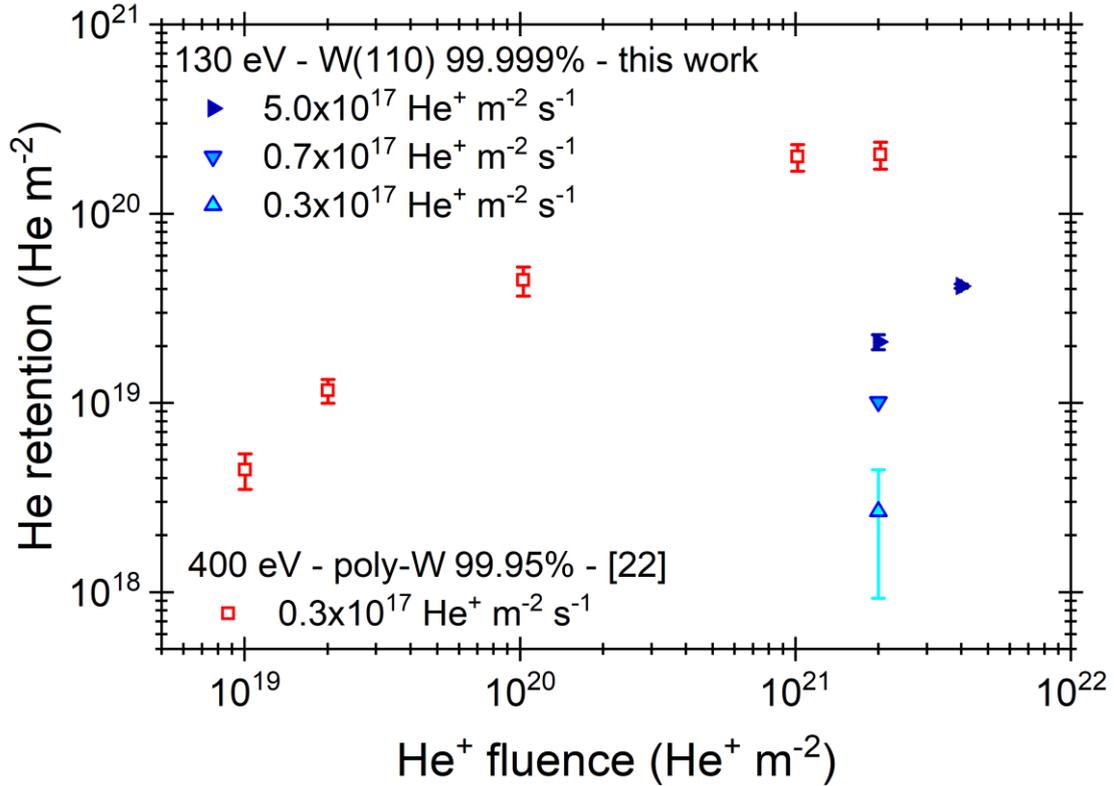

Figure 4. He retention as a function of He$^+$ fluence implanted at room temperature for ion kinetic energy below the knock-on threshold for displacement damage in W. Blue triangles represent this work's *in situ* TPD measurement performed in clean single crystal W(110) with 99.999% purity, for various implantation flux. The open red squares corresponds to *ex situ* TPD measurements realized by Gasparyan *et al.* in polycrystalline W with 99.95% purity [22] with a He$^+$ flux below the self-trapping threshold determined in this work.

Once the flux threshold for He self-trapping is overcame, our measured He retention for a fluence of $2.0 \times 10^{21}$ He$^+$.m$^{-2}$ is increasing and at a flux of $5.0 \times 10^{17}$ He$^+$.m$^{-2}$.s$^{-1}$ the present He retention is an order of magnitude smaller than the one obtained by Gasparyan *et al.*. For the same flux and fluence, when we increase the kinetic energy to 620 eV to introduce knock-on displacement damage during implantation, the He retention further increases by a factor of four (not shown) and only a factor of 2.5 separates our He retention at $2.0 \times 10^{21}$ He$^+$.m$^{-2}$ to the one of Gasparyan *et al.*. These comparisons highlight that He retention is a complex function of both the He flux and the concentration of preexisting defects. The last experiment shown in Figure 4, with the increased He retention measured for an increased fluence of $4.0 \times 10^{21}$ He$^+$.m$^{-2}$ and a flux of $5.0 \times 10^{17}$ He$^+$.m$^{-2}$.s$^{-1}$, confirms that He retention is also a function of the fluence as shown by Gasparyan *et al.*.

## 4. Conclusions and perspective

In the present study, we prepared a pure W(110) sample with a clean surface in order to evidence the self-trapping of He in the W bulk at a sample temperature of 300 K and for a constant fluence of $2.0 \times 10^{21}$

He$^+$.m$^{-2}$. At a He$^+$ kinetic energy of 130 eV and a flux of $0.3 \times 10^{17}$ He$^+$.m$^{-2}$.s$^{-1}$, we only observed a small He desorption peak below 600 K which is consistent with the dissociation energy of mobile He$_n$ interstitial clusters as determined by DFT by Boisse *et al.* [17]. Rising the ion flux to $0.7 \times 10^{17}$ He$^+$.m$^{-2}$.s$^{-1}$, we observed the sudden appearance of two desorption peaks at 950 K and 1700 K which temperature positions are consistent with the dissociation energy of self-trapped He$_n$V clusters [17,32]. For the highest flux studied in this work, $5.0 \times 10^{17}$ He$^+$.m$^{-2}$.s$^{-1}$, an additional desorption peak at 1800 K and a desorption shoulder at 1900 K are observed, consistently with the occurrence of trap-mutated He$_n$V$_2$ and He$_n$V$_3$ clusters [17]. These experimental observations prove that the He$^+$ flux has a threshold effect on He retention and He self-trapping and that, for a constant temperature and fluence, increasing the flux increases the number of type of self-trapped and trap-mutated He$_n$V$_m$ clusters in the W bulk. We have also confirmed that preexistent vacancy and vacancy clusters, as created e.g. with knock-on displacement damage, give birth to two additional desorption peaks: an intense and broad peak around 1100 K that should be related to He$_n$V$_m$ vacancy clusters and a small peak around 640 K that may be related to near-surface He$_n$V$_m$ clusters.

The present set of results should be used to constrain the development of He retention and He bubbles growth models based on *ab initio* quantities, e.g. such as the ones developed by Faney *et al.* [18] and Delaporte-Mathurin *et al.* [19]. In future works, we plan to extend the sample temperature range for which He$^+$ implantation is realized, to determine the temperature dependence of the He$^+$ flux threshold. Previous modeling work by Boisse *et al.* [17], Faney *et al.* [18] and Delaporte-Mathurin *et al.* [19] have shown such temperature dependency on He self-trapping indeed. Finally, the effect of surface contamination on He retention highlighted by Kornelsen [12] and Gaspayan *et al.* [22] should be explored systematically, especially trying to decipher the effect of surface contamination on both interstitial He$_n$ clusters and self-trapped He$_n$V clusters.

### Acknowledgments

This work has been carried out within the framework of the French Federation for Magnetic Fusion Studies (FR-FCM). The project leading to this publication has received funding from the Excellence Initiative of Aix-Marseille University—A∗Midex, a French 'Investissements d'Avenir' programme as well as from the ANR under Grants ANR-18-CE05-0012 and ANR-20-CE08-0003.